\documentclass[
    ,final            
  ]
  {aipproc}

\layoutstyle{8x11single}
\usepackage{amssymb}

\begin{document}

\title{Dark Matter and Dark Energy \\
from Gravitational Symmetry Breaking}

\classification{<Replace this text with PACS numbers; choose from this list:
                \texttt{http://www.aip..org/pacs/index.html}>}

\author{A. F\"uzfa}{
  address={Department of Mathematics, University of Namur (FUNDP); Rue de Bruxelles, 61, B-5000 Namur, Belgium},
  altaddress={Center for Particle Physics and Phenomenology (CP3), Universit\'e catholique de Louvain, Chemin du Cyclotron, 2, B-1348 Louvain-la-Neuve, Belgium}
}

\author{J.-M. Alimi}{
 address={CNRS, Laboratoire Univers et Th\'eories (LUTh), UMR 8102 CNRS, Observatoire de Paris, Universit\'e Paris Diderot ; 5 Place Jules Janssen, 92190 Meudon, France}}

\begin{abstract}
We build a mechanism of gravitational symmetry breaking (GSB) of a global $U(1)$ symmetry 
based on the relaxation of the equivalence principle due to the mass variation of pseudo Nambu-Goldstone
dark matter (DM) particles.
This GSB process
is described by the modified cosmological convergence mechanism of the Abnormally Weighting Energy (AWE) Hypothesis previously
introduced by the authors. Several remarkable constraints from the Hubble diagram of far-away supernovae are derived,
notably on the explicit and gravitational symmetry breaking energy scales of the model. We then briefly present some consequences
on neutrino masses when this mechanism is applied to the particular case of the breaking of lepton number symmetry.
\end{abstract}

\maketitle


\section{Introduction}

Symmetries induspitably play a central role in physics. They not only constitute a powerful tool to build 
efficient and smart fundamental laws, they also allow us identifying the deep nature of physical interactions
and the properties of matter and space-time. In many situations however, this is not exact symmetry
that is sought for to give the richness of the physical description but rather approximate or broken
symmetries. Indeed, a broken symmetry lies behind several crucial questions in modern physics, including
for instance
the mass generation of elementary particles through the Higgs-Brout-Englert mechanism, the cleaning up of the strong-CP problem in
quantum chromodynamics or the observed non vanishing neutrino masses.
\\
\\
By investigating the physical mechanisms that could be responsible for the breaking of these fundamental symmetries,
it appeared several other implications that are of first interest for cosmology. A famous example is the
existence of a light pseudo-scalar particle, the $axion$, predicted by Weinberg and Wilczek \cite{axion}, as
a by-product of the breaking of the Peccei-Quinn symmetry that could cure the strong-CP problem of QCD \cite{PQ}. Such 
light particles, interacting very weakly with the other particles in the standard model, then constitute
ideal candidates for solving another key question of modern cosmology: the nature of Dark Matter (DM).
Other particles similar to the Weinberg-Wilczek axion, whose existence are motivated by other problems in particle physics,
could also constitute good particle physics for at least a part of the enigmatic DM. One famous example is the majoron \cite{majoron}, which
appears in the breaking of the global lepton number symmetry in the standard model.
\\
\\
One key property of such axions, that has strong implications on the way they could possibly solve
problems in particle physics $and$ in cosmology, is of course their mass. This question is obviously closely related
to the particular mechanism of symmetry breaking. In the case of the Weinberg-Wilczek axion, the global $U(1)_{PQ}$
Peccei-Quinn symmetry is spontaneously broken at some high energy scale when a new scalar field, charged under
this symmetry, acquires some vacuum expectation value (vev). Then the couplings between ordinary matter particles
and the new scalar field yield an effective mass-term for the Nambu-Goldstone boson associated to the broken symmetry.
This results in an explicit symmetry breaking (ESB) of the global $U(1)_{PQ}$ symmetry. It is important to notice
that this ESB is usually considered of perturbative nature and that the
effective mass for the axion is temperature dependent due to the running of its couplings to matter during
the whole process of symmetry breaking \cite{kolb}. 
\\
\\
Therefore, the axion really constitutes the prototype of \textit{mass-varying} DM. The particular
interactions the DM particle could exert with ordinary matter particules have been long time identified as an important
feature to track it down through non standard gravitational effects on cosmological scales. Several models based on  particle physics have been considered to achieve such a mass-variation for DM (cf. \cite{farrar,massd,mavans,dmde} and references therein). The consequence of DM mass variation is that it therefore violates
the weak equivalence principle (WEP), that establishes the universality of free-fall for non-gravitationally bound objects.
This means that, unlike in General Relativity (GR), gravitation is sensitive to the various nature and composition of the test bodies,
particularly if those are made of mass-varying DM.
While the universality of free fall has been extremely well-verified, notably at the $10^{-12}$ level with laboratory masses \cite{su},  these tests exclusively hold for ordinary matter \textit{only}.
The question whether the WEP can be applied or not to DM still remains an open debate, on both observational \cite{kamionkowski} and theoretical \cite{farrar,massd,mavans,dmde} points of view.
\\
\\
In addition, the fact that DM could interact very weakly with other matter components also raises the question
of its possible relation with another enigmatic feature of the large-scale universe:
cosmic acceleration. Cosmic acceleration is usually assumed to be related
to the existence of a new energy component, Dark Energy (DE), that acts as a repulsive gravitational force on cosmic
expansion through important negative pressures. In the convenient picture of
the concordance model $\Lambda \rm CDM$, DE is related to a non vanishing cosmological constant and DM is a yet 
undiscovered new heavy (and long-lived) particle.
The dark sector, which accounts for 96\% of the total energy content of the Universe is usually assumed
to be constituted of these two ingredients whose physical nature is completely different. 
In the same time, 
there are several physical motivations to go beyond this simplistic view by considering 
some more sophisticated physics in the dark sector (see e.g. \cite{farrar,dmde,awe}), for instance by introducing interactions
between DM and DE. Such interactions between DM and DE are long-ranged, due to the lightness of the DE field, which
introduces a whole new phenomenology of the invisible sector on cosmological scales.
Such phenomenology of DM particles with a mass ruled by the DE scalar field has been extensively
studied. Several approaches and motivations to mass-varying DM can be found in \cite{farrar,massd}.
Other authors \cite{mavans} have developped interesting interpretations of the dark sector based on neutrino physics by coupling them
to the DE scalar field which makes the small mass of the neutrinos varying along cosmic history.
\\
\\
The physics of DM, and quite possibly those of DE if they turn out to be only different aspects of the same physical process,
really question the validity of the equivalence principle on cosmological scales where DM is profuse. Yet the equivalence principle
is one of the most precisely verified hypothesis in fundamental physics \cite{will}, for what
concerns ordinary matter and sub-galactic scales. Then, if it turn out to be violated by DM or
the dark sector of cosmology, the central question is: how far is ordinary matter \textit{aware}
of what is going wrong with the equivalence principle in the dark sector? As we do not have in hand physical tools
made of DM, it is impossible to reveal such violation directly on DM itself. Is there any way ordinary matter
could experience \textit{directly} the violation of the EP by the dark sector? The answer is certainly yes, actually through
two distinct modified gravity effects induced by mass-varying DM. 
The first, introduced by Carroll and his collaborators \cite{mantry} is a very faint violation of the 
WEP \textit{in the visible sector}. This effect is small, since it is related to high-order loop corrections, but it
could be detected in future
experiments aiming at verifying universality of free fall for ordinary matter particles with very high accuracy. The second effect on usual particles
due to the abundance of mass-varying DM is an induced violation of the \textit{strong} equivalence principle (SEP),
introduced by Alimi \& F\"uzfa \cite{awe}.
The SEP is an extension of the WEP to gravitational binding energies, yielding that gravitation is produced in the same way
by any kind of energy, including gravitational ones. In the presence of mass-varying DM, this principle does not hold anymore:
some gravitationally bound test body made of ordinary and DM will weigh differently throughout space-time 
accordingly with the variation of DM inertial mass. Such mass-varying DM particles therefore constitute an Abnormally Weighting type of Energy (AWE Hypothesis). In other words, mass-varying DM induces a change in the gravitational strength
experienced by ordinary matter. This effect can be expected to be particularly large when DM is profuse,
especially in the large-scale universe.
\\
\\
To summary, the questions of symmetry breaking and hidden interactions in the dark sector quite naturally lead
to question the EPs and the deep nature of gravitation. However, the EPs have
been precisely tested for ordinary matter on small scales. Therefore, if the EPs are to be modified
considering the above-mentionned problems, there is only room for a relaxation of the EPs on cosmological scales.
Still such a modification requires a covariant theory of gravitation that describes completely all the aspects of EP
relaxation due to the abundance of mass-varying DM: violation of both WEP and SEP. 
This has been achieved in \cite{awe} through the widespread tool of tensor-scalar theories (TST) of gravitation.
One remarkable property of such theories is that they incorporate a natural establishment of the EPs during cosmological expansion  \cite{convts,serna,awe}, through the attraction of the extra gravitational scalar degrees of freedom toward GR-like states.
\\
\\
It has been long time conjectured that gravitation should play a role in the breaking of global U(1) symmetries \cite{majoron}.
However, this argument has always been translated so far in terms of the scale at which the spontaneous symmetry breaking (SSB)
occurs: a scale typically comparable to the Planck mass or at least much higher than the electroweak scale. 
We introduce here a new mechanism that achieves this idea that gravitation
somehow rules the way scalar fields charged under global symmetries are stabilized. The mechanism we propose here is not based
anymore on some Nambu-Goldstone potential but rather makes use of the cosmological attraction mechanisms to stabilize
the vev of those scalar fields. The non-minimal gravitational coupling of such fields plays a central role in this new
mechanism. To be more precise, our argument is as follows. Since the symmetry breaking process makes the mass running, this
yields a violation of the WEP and consequently a variation
of the gravitational strength (SEP violation). The dynamics of the gravitational strength is then attracted toward
some values during cosmic expansion, which allows a classic and purely gravitational stabilisation of the scalar fields vev.
Therefore, gravitation is substantially modified during the stabilisation process which results in a non-standard
Friedmann-Lemaître cosmic expansion. In particular, cosmic acceleration is produced which relates naturally symmetry breaking
process, mass-varying DM and the last component of the dark sector: DE.
\\
\\
The structure of this paper is as follows. In the next section, we build the gravitational symmetry breaking (GSB) mechanism 
of a global $U(1)$ symmetry before we present the cosmological evolution and the possibility to reproduce
cosmic acceleration during the GSB process. We then focus on cosmological constraints
of this mechanism from the Hubble diagram of the UNION data set of type Ia supernovae data (SNe Ia) \cite{union}. This allows us discussing some physical predictions of this mechanism
in the case where the choice of the global symmetry is motivated by neutrino physics. We conclude in the last section by some perspectives.
\section{An effective action for gravitational symmetry breaking (GSB)}
Problems like the weakness of the neutron electric dipole moment or the small yet non vanishing neutrino mass
can be solved through the explicit symmetry breaking (ESB) of a global abelian symmetry ($U(1)_{PQ}$ Peccei-Quinn symmetry
for the strong-CP problem ; $U(1)_{B-L}$ lepton number symmetry for neutrino masses). The mechanism is usually as follows.
One introduces a new complex scalar field $\Psi$ to the standard model, a field that is charged under the global $U(1)$ and coupled
to some of the matter fields (the neutrinos for instance in the case of $U(1)_{B-L}$). If the field $\Psi$ acquires some vev, the global symmetry is spontaneously broken, which is usually
done through the introduction of a Nambu-Goldstone potential:
\begin{equation}
V_{NG}(|\Psi|)\approx\left(|\Psi|^2-\lambda_S^2\right)^2
\label{vnambu}
\end{equation}
with $\lambda_S$ the SSB scale. Around the minimum of this mexican hat potential, we can write
down
\begin{equation}
\Psi=\lambda_S e^{i\frac{\theta(x^\mu)}{\lambda_S}}
\end{equation}
with $\theta$ is the massless Goldstone boson associated to the symmetry breaking. However, the couplings between
$\Psi$ and other matter fields give rise to ESB  through the constitution of an effective
mass term for the Goldstone boson $\theta$. Such mass term varies in general during the phase transition \cite{kolb}
and the massive scalar field is dubbed the axion. Because of the residual $U(1)$ symmetry after
the SSB, the effective potential giving the mass to $\theta$ must be periodic (residual discrete symmetry) so that the mass of the axion $\theta$ is
protected from quantum fluctuations. Such effective potential can be written as in \cite{pngb}:
\begin{equation}
V_{\rm eff}(\theta)=\lambda_e^4 \left(1+\cos\left(\frac{\theta(x^\mu)}{\lambda_S}\right)\right)\cdot
\label{vfrieman}
\end{equation}
The previous potential has been widely used in the litterature and constitutes the pseudo-Nambu-Goldstone boson (pNGB) candidate
for quintessence \cite{pngb}. Indeed, when the scalar field is frozen on top of the potential, it acts
as a cosmological constant whose density is related to the ESB scale by $\rho_{\Lambda}\approx \lambda_e^4$.
This is why the most plausible candidate for this pNGB comes from neutrino physics since the observed
value $\rho_{\Lambda}\approx (10^{-3}eV)^4$ lies in the vicinity of the neutrino mass discrepancies. When the field $\theta$
falls down around the minimum of (\ref{vfrieman}), it acquires a mass and classicaly behaves as some cold DM (see for instance
\cite{das} which uses this feature for early DM generation).\\
\\
Such complex scalar field are particularly appealing for trying to solve DM and DE problems at the same time. Indeed, one can
identify $\sigma=|\Psi|$ with the DE scalar field and $\theta=Arg(\Psi)$ to axion-like DM. Several attempts
have dealt with this idea \cite{cplxdmde}, for instance by simply replacing the Nambu-Goldstone potential
(\ref{vnambu}) by another shape of potential more appropriate for explaining the late cosmic acceleration
 (e.g. tracking or run-away potentials). However, this replacement looks rather arbitrary and ad hoc, if not rather incompatible
with constraints from microphysics.\\
\\
Let us now build a gravitational symmetry breaking mechanism for the complex scalar $\Psi$ charged under some general $U(1)$
 global symmetry. We first consider the lagrangian of the complex scalar
 itself which we write down
\begin{equation}
\mathcal{L}_{\Psi}=\partial_\mu\sigma\partial^\mu\sigma+\frac{\sigma^2}{\lambda_g^2}\partial_\mu\theta\partial^\mu\theta
+\lambda_e^4\left(\frac{\sigma}{\lambda_g}\right)^4 \left(1+\cos\left(\frac{\theta}{\lambda_g}\right)\right)
\label{lpsi}
\end{equation}
 where we set
\begin{equation}
\Psi=\sigma e^{i\frac{\theta}{\lambda_g}}
\end{equation}
with $\lambda_g$ an energy scale to be determined further. The effective mass of the axion-like particle $\theta$
is driven by the field $\sigma$: 
$$
m_{\theta}=\frac{\lambda_e^2}{\lambda_g}M(\sigma)\equiv \bar m_{\theta}M(\sigma)
$$
with $\bar m_{\theta}$ is the bare mass of the pNGB and $M(\sigma)$ is a mass scaling to be specified further. 
The vev of $\sigma$ has still to be stabilised, which will be the goal of the GSB mechanism.\\
\\
When the field $\theta$ oscillates in the bottom of the potential term in (\ref{lpsi}), it appears as
a DM particle with varying mass (given by $M(\sigma)$). Therefore, when coupling the dynamics of $\Psi$ to gravitation, we cannot
use the standard Einstein-Hilbert action. Indeed, the EPs have to be relaxed due to the mass variation of this species
and we will write down the purely gravitational part of the action as a Brans-Dicke-like lagrangian:
 \begin{equation}
 \mathcal{L}_{\rm grav}=\Phi(\sigma) R-\frac{\Omega(\Phi)}{\Phi}\partial_\mu\Phi\partial^\mu\Phi
\label{lgrav}
 \end{equation}
where $\Phi$ is a scalar degree of freedom scaling the gravitational strength in the theory,
$R$ is the Ricci scalar build upon the metric $g_{\mu\nu}$
and $\Omega(\Phi)$ is some arbitrary function of $\Phi$. Since we know
that the cause of this SEP violation stands in the symmetry breaking, the gravitational
strength depends on the vev of $\Psi$ and we have
$\Phi=\Phi(\sigma)$. In other words, the vev of the complex scalar rescales both the gravitational strength and the pNGB mass.\\
\\
Finally, we have to introduce ordinary matter in this picture. If the ordinary matter fields are not charged under the
global $U(1)$ symmetry (like for the lepton number symmetry), they do not couple directly to $\Psi$ and we have that
the matter lagrangian does not explicitely depend on $\sigma$ and therefore on $\Phi$. As well, this assumption is also
valid if the couplings between matter fields and $\Psi$ are weak. As we shall see below, this allows a retricted WEP to hold
for the ordinary matter sector.\\
\\
We can now combine (\ref{lpsi}) and (\ref{lgrav}) and focus only on the field $\Phi$, since $\Phi$ and $\sigma$
are directly related. Taking into account that the ordinary matter fields do not couple directly\footnote{Or very weakly}
to $\Psi$, we have the following action for the GSB mechanism
($c=1$ ; $\kappa=8\pi/m_{Pl}^2$ with $m_{Pl}$ the Planck mass): 
\begin{eqnarray}
\label{gsb}
S_{\rm GSB}&=&\frac{1}{2\kappa}\int\;\sqrt{-g}d^4 x\left\{\Phi R-\frac{\omega(\Phi)}{\Phi} \partial_\mu\Phi\partial^\mu \Phi\right\}\nonumber\\
&&+S_{\theta}\left[\theta,M^2(\Phi)g_{\mu\nu}\right]+S_m\left[\psi_m,g_{\mu\nu}\right]
\end{eqnarray}
where $g_{\mu\nu}$ is actually the metric coupling universally to ordinary matter (Jordan frame metric), $\omega(\Phi)$ is the Brans-Dicke coupling function while $\psi_{m}$
are the fundamental fields entering the physical description of the ordinary matter sector. The pNGB field $\theta$ abnormally
weighs: it does not fall along the geodesics of $g_{\mu\nu}$, the metric field that
is measured through rods and clocks made of ordinary matter, but rather along the geodesics of $\bar g_{\mu\nu}=M^2(\Phi)g_{\mu\nu}$.
This is the mathematical translation of the violation of the WEP by the abnormally weighting energy of $\theta$.
The function $M(\Phi)=\sigma(\Phi)/\lambda_g$ represents the mass-variation of the pNGB $\theta$ during the symmetry breaking process, and is represented as a non-minimal coupling to the metric $g_{\mu\nu}$.
However, the local laws of physics for ordinary matter are those of special relativity (as the matter action $S_m$ does not explicitely depend on the 
scalar field $\Phi$). A restricted version of the WEP therefore applies to the ordinary matter sector: we have only relaxed the WEP
by assuming that only species, the pNGB field $\theta$, is a gravitational outlaw. 
\\
\\
The Einstein field equations that follow from Eq. (\ref{gsb}) are almost the same than those of the usual tensor-scalar case (\cite{convts,serna}), but in this case the source term is
the sum of $\theta$ and ordinary matter contributions $\left(T_{\mu\nu}^m+T_{\mu\nu}^{\theta}\right)$. 
The Klein-Gordon equation for the scalar field $\Phi$, however, is modified due to the violation of the WEP and writes down
\begin{eqnarray}
\label{kg}
\left(3+2\omega(\Phi)\right)\Box\Phi&=&\kappa \left\{T^m+\left(1-2\Phi \frac{d\ln M(\Phi)}{d\Phi}\right)T^{\theta}\right\}
-\frac{d\omega(\Phi)}{d\Phi}g^{\alpha\beta}\partial_{\alpha}\Phi\partial_{\beta}\Phi,
\end{eqnarray}
where $T^i=g^{\mu\nu}T_{\mu\nu}^i$ stands for the trace of the stress-energy tensor for species $i$. As the WEP is verified for ordinary matter, its conservation equations lie
unchanged, $\nabla_\mu T_\nu^{m\;\mu}=0$, while it is not the case for the abnormally weighting sector: 
\begin{equation}
\label{cons3}
\nabla_\mu T_\nu^{\;\mu}(\theta)=\frac{d\ln M(\Phi)}{d\Phi}\;T^{\theta}\partial_\nu\Phi \cdot
\end{equation}
These equations hold for the Jordan frame, which is more intuitive to express the observable quantities. A similar description
of the model in terms of Einstein frame variables can be found in \cite{awe}
and \cite{alimi}.
\section{The cosmological mechanism of GSB}
With these tools in hand, we can now describe the cosmological dynamics for the GSB
mechanism.
We assume 
a flat Friedmann-Lema\^itre-Robertson-Walker (FLRW) metric: 
\begin{equation}
ds^2=-dt^2+a^2(t)dl^2\cdot
\label{flrw}
\end{equation} 
and write down the Friedmann, acceleration and Klein-Gordon equations ($M(\Phi)$ and $\omega(\Phi)$ will be specified further):
\begin{eqnarray}
\left(\frac{\dot{a}}{a}\right)^2&=&-\frac{\dot{a}}{a}\frac{\dot{\Phi}}{\Phi}+\frac{\omega}{6}\frac{\dot{\Phi}^2}{\Phi^2}+\frac{\kappa}{3\Phi}\left(\rho_m+\rho_{\theta}\right)\label{fried}\\
2\frac{\ddot{a}}{a}&=&-\frac{\dot{a}}{a}\frac{\dot{\Phi}}{\Phi}-\frac{\ddot{\Phi}}{\Phi}-\frac{2\omega}{3}\left(\frac{\dot{\Phi}}{\Phi}\right)^2\nonumber\\
&& -\frac{\kappa}{3\Phi}\left(\rho_m+3p_m+\rho_{\theta}+3p_{\theta}\right)\label{acc}\\
\left(3+2\omega\right)\left(\ddot{\Phi}+3\frac{\dot{a}}{a}\dot{\Phi}\right)&=&\kappa\left(\rho_m-3p_m+\left(\rho_{\theta}-3p_{\theta}\right)\left(1-2\Phi \frac{d\ln M(\Phi)}{d\Phi}\right)\right)-\frac{d\omega}{d\Phi}\dot{\Phi}^2\label{kgcos}\\
\ddot{\theta}+3\frac{\dot{a}}{a}\dot{\theta}&=&-2\dot{\theta}\dot{\Phi}\frac{d\ln M(\Phi)}{d\Phi}-\frac{M^2(\Phi)}{2}\frac{dV}{d\theta}\label{eqth}
\end{eqnarray}
where a dot denotes a derivative with respect to observable cosmic time and where
the quantities $\rho_\theta$, $p_\theta$ and $V$ stands for the energy density, pressure and effective potential, respectively.
These are given by
\begin{eqnarray}
\rho_\theta&=&M^2(\Phi)\left(\dot{\theta}^2+M^2(\Phi)V(\theta)\right)\\
p_\theta&=&M^2(\Phi)\left(\dot{\theta}^2-M^2(\Phi)V(\theta)\right)\\
V(\theta)&=&\lambda_e^4 \left(1+\cos\left(\frac{\theta}{\lambda_g}\right)\right)\label{pot}\cdot
\end{eqnarray}
\\
\\
We can partly decouple the equations for the scalar fields (\ref{kgcos}) and (\ref{eqth}) by defining
(cf. \cite{serna})
\begin{eqnarray}
\psi&=&\frac{1}{2}\ln\Phi\nonumber\\
3+2\omega&=&3W\\
h&=&\frac{dN}{dt}=\frac{\dot{a}}{a}+\dot{\psi}\cdot\nonumber
\end{eqnarray}
with $N$ the number of efoldings of the Einstein frame metric $g^*_{\mu\nu}=\Phi g_{\mu\nu}$. 
Eqs. (\ref{fried}) and (\ref{acc}) now reduce to 
\begin{eqnarray}
3h^2\left(1-W\psi^{'2}\right)=\kappa e^{-2\psi}\left(\rho_m+\rho_{\theta}\right)\\
\frac{h'}{h}=-\left(1-\psi'\right)-2W\psi^{'2}-\frac{\left(1-W\psi^{'2}\right)}{2}\left(1+3\lambda_T\right)
\end{eqnarray}
where a prime denotes a derivative with respect to $N$ and $\lambda_T=(p_m+p_{\theta})/(\rho_m+\rho_{\theta})$ is the total equation of state for the background fluid
composed of ordinary matter and abnormally weighting field $\theta$.\\
\\
Eqs (\ref{kgcos}) and (\ref{eqth}) now become
\begin{eqnarray}
\label{dyn}
\frac{2\psi''}{1-W\psi^{'2}}+3\left(1-\lambda_T\right)\psi'&=&\frac{1}{W}\left(1-3\lambda_T-\frac{d\log M}{d\psi}\left(1-3\lambda_{\theta}\right)\frac{\rho_{\theta}}{\rho_{\theta}+\rho_m}\right)\nonumber\\
&&-\frac{\psi^{'2}}{W(1-W\psi^{'2})}\frac{dW}{d\psi}\\
\theta''+\frac{3}{2}\left(1-W\psi^{'2}\right)\left(1-\lambda_T\right)\theta'&=&2\psi'\theta'\left(1-\frac{d\log M}{d\psi}\right)-\frac{M^2}{2h^2}\frac{dV}{d\theta}
\label{dyn2}
\end{eqnarray}
where $\lambda_\theta=p_\theta/\rho_\theta$.
The term proportional to $1-3\lambda_{\theta}$ in (\ref{dyn}) accounts for the relaxation of the WEP and
can be neglected if $\rho_{\theta}\ll \rho_m$.
This is achieved during the radiation-dominated era when the field $\theta$ slowly rolls on top of the potential (\ref{pot}) 
($\rho_\theta\approx \rm cst$ and $\rho_m\approx a^{-4}$), therefore the
other scalar field\footnote{Not to be confused with the complex scalar $\Psi$.} $\psi$ evolves like in usual TST: it is quickly damped by cosmic expansion toward GR ($\Phi\approx 1$ $M(\Phi)\approx 1$). 
The scalar field $\theta$ falls into the bottom of the potential (\ref{pot}) when 
$m_\theta=\lambda_e^2/\lambda_g\approx 3 H$, which corresponds to the transformation of $\theta$ in CDM at epoch 
$a_{\rm CDM}\approx \left(3\sqrt{2}\frac{\lambda_g}{\lambda_e^2}\left(\frac{\kappa}{3}\rho_{\rm rad}(a_0)\right)^{1/2}\right)^{1/2}$.
Therefore, the ratio $R_i$ between the amount of axion-like DM ($\theta$) and baryons (ordinary pressureless matter)
is fixed by the scales $\lambda_g$ and
$\lambda_e$.\\
\\
When we enter matter-dominated era, the scalar field $\psi$ unfreezes and its dynamics is given by (\ref{dyn}) with $\lambda_T\approx\lambda_\theta\approx 0$. 
Eq. (\ref{eqth}) (or equivalently the conservation equation for the AWE component) can be solved
$\rho_{\theta}=C_{\theta} \Phi a^{-3}$, with some constant $C_\theta$.
In the matter-dominated era, there now exists two fixed points $\psi_\infty$ for the field $\psi$ given by either
$W(\psi_\infty)=\infty$, which corresponds to GR, and 
$$
\frac{d\log M}{d\psi}|_{\psi=\psi_\infty}=1+R_i e^{-2\psi_\infty},
$$
where $R_i$ is the ratio at which the energy densities of baryons and pNGB $\theta$ DM froze during radiation-dominated era.
For instance, if we choose $M(\Phi)=\Phi$, this last fixed point corresponds to:
\begin{equation}
\psi_\infty=\frac{1}{2}\ln\left(R_i\right)\cdot
\end{equation}
The second
fixed point corresponds to a GR-like theory but with a value of the gravitational strength different from Newton's constant. 
A stability
analysis shows (cf. \cite{awe}) that the first fixed point with the gravitational strength equal to Newton's constant (GR) is
unstable while the other is stable when $DM$ is profuse ($\rho_m\ll\rho_\theta$). In the opposite situation of small
influence of DM ($\rho_m\gg\rho_\theta$), there is only one attractor (a fixed point that is stable) corresponding to GR. \\
\\
Therefore, a substantial departure from standard FLRW cosmic expansion occurs during the cosmological transition between $\Phi(a\approx 0.001)=1$ (GR) at the end of radiation-dominated
era and $\Phi_\infty=R_i$. This effect can be observed as cosmic acceleration as we shall see below
\section{Cosmic acceleration as a consequence of GSB}
Using Friedmann (\ref{fried}) and Klein-Gordon (\ref{kgcos}) equations, we can rewrite the acceleration equation (\ref{acc}) during the matter-dominated era 
($\lambda_m\approx\lambda_{\theta}\approx 0$):
\begin{eqnarray}
\frac{\ddot{a}}{a}&=&\overbrace{\frac{\dot{a}}{a}\frac{\dot{\Phi}}{\Phi}-\frac{\omega}{3}\frac{\dot{\Phi}^2}{\Phi^2}-\frac{1}{2}\left(3+2\omega\right)^{-1}\frac{\dot{\Phi}^2}{\Phi}\frac{d\omega}{d\Phi}}^{I}\nonumber\\
&&\underbrace{-\frac{4\pi}{3}G_m^{\textrm{eff}}\rho_m}_{II}+\underbrace{\frac{4\pi}{3}G_{\theta}^{\textrm{eff}}\rho_{\theta}}_{III},\label{acc2}
\end{eqnarray}
where there appears two distinct cosmological Cavendish couplings for the visible and invisible sectors:
\begin{eqnarray} 
G_{m}^{\textrm{eff}}&=&\frac{G}{\Phi}\left(\frac{6+2\omega}{3+2\omega}\right)\\
G_{\theta}^{\textrm{eff}}&=&\frac{G}{\Phi}\left(\frac{2\omega}{3+2\omega}\right)
\end{eqnarray}
where we have assumed for $M(\Phi)=\Phi$ for simplicity\footnote{For the numerical results below, we also chose $3+2\omega(\Phi)=1/(K\ln\Phi)$ with $K$ 
some parameter but the results presented here do not depend on the particular shape of this coupling function, provided
 $3+2\omega\ge 0$ and $\omega=\omega(\Phi)$ (cf. \cite{awe}). }.
The cosmological Cavendish couplings of pNGB $\theta$ can be negative for appropriate values of $\Phi$ while the one of ordinary
matter will always be positive. Therefore, the mass variation of the $\theta$ particle that causes a relaxation of WEP can result in a repulsive gravitational force which will produce cosmic acceleration according to (\ref{acc}) and (\ref{acc2}), provided the amount of CDM is predominant ($\rho_m\ll\rho_{\theta}$). 
It is very important to note that this repulsive action is achieved without appealing to a ill-defined negative gravitational coupling $G_c$ at the level of the action (\ref{gsb}), as we always have $0\le\Phi\le 1$.


To summary, the cosmological evolution during the GSB is as follows (see also \cite{awe}).
As the Universe cools down in radiation-dominated era, the amount of ordinary matter and pNGB DM $\theta$ becomes comparable
for appropriate values of the scales $\lambda_g$ and $\lambda_e$. These scales fix the coincidence problem between baryons and CDM.
The repulsion due to the WEP relaxation progressively switches on as the influence of DM energy density
increases. It will then finally dominate the cosmological dynamics after matter-radiation decoupling. Starting near GR ($\psi\rightarrow 0$ and $W\rightarrow\infty$) at the beginning of matter-dominated era, the scalar field $\Phi$
starts deviating from this unstable state where pNGB DM $\theta$ and ordinary matter weigh differently and is then attracted toward the GR-like state $\psi_{\infty}$ where the ordinary matter and AWE equally weigh: $\rho_{\theta}\rightarrow \rho_m$ (for $M(\Phi)=\Phi$). During the establishment of this new gravitational equilibrium, several phases of cosmic acceleration are achieved. It is then possible to reproduce the Hubble diagram of type Ia supernovae
with this cosmological GSB mechanism.\\
\\

\begin{figure}[h!t]
  \includegraphics[scale=0.5]{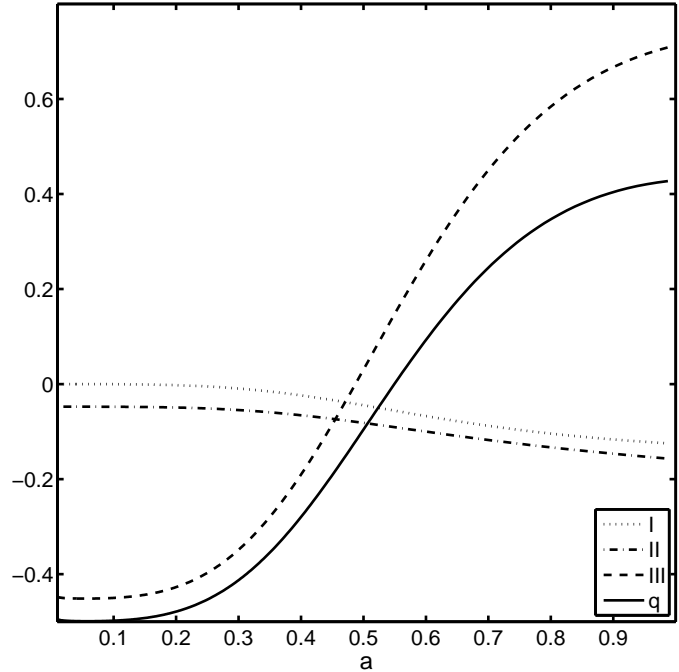}
  \caption{Acceleration factor $q=\ddot{a}a/\dot{a}^2$ and its constitutive terms from Eq.(\ref{acc2}). Cosmic acceleration
is produced by the repulsive influence of abnormally weighting DM field $\theta$}
\label{fig_acc}
\end{figure}

\begin{figure}[h!t]
  \includegraphics[scale=0.5]{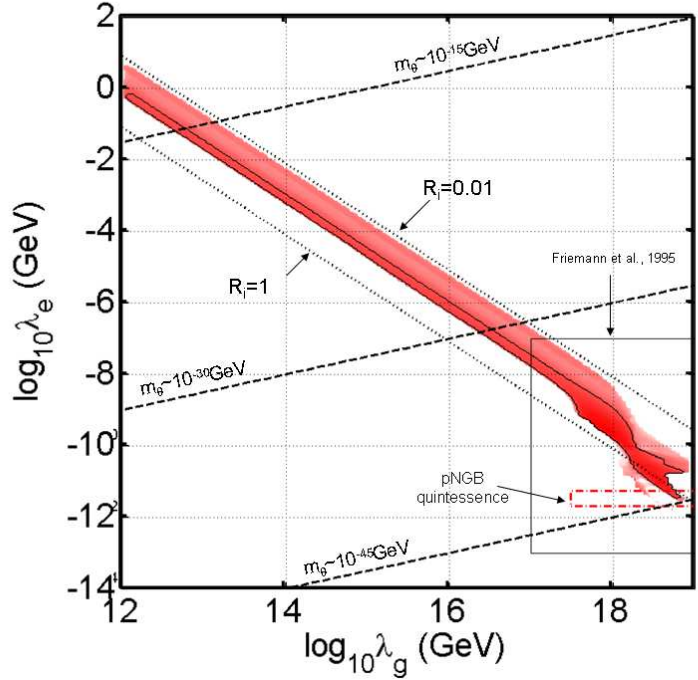}
  \caption{Constraints from UNION supernovae data on the energy scales $\lambda_e$ and $\lambda_g$ of
the GSB mechanism}
\label{lgle}
\end{figure}

Figure \ref{fig_acc} illustrates the evolution of the acceleration factor $q=\ddot{a}a/\dot{a}^2$
from Eq.(\ref{acc2}) in a model that fits UNION supernovae data \cite{union} with the same accuracy as the concordance
model $\Lambda$CDM. 
 Also represented in Figure \ref{fig_acc} are the contributions to it from the different terms of Eq.(\ref{acc2}). Cosmic acceleration is produced
through the last term in Eq.(\ref{acc2}) which causes a repulsive action from the negative cosmological Cavendish coupling
of abnormally weighting field $\theta$. The model upon which Figure \ref{fig_acc}
has been build corresponds to the following values of the energy scales: $\lambda_e\approx 1MeV$
$\lambda_g\approx 10^{14}GeV$
which correspond to $R_i=0.1$, $\Omega_b=0.04$, $\Omega_\theta=0.25$ ($\Omega_\Phi=0.69$) and $\bar m_\theta\approx 10^{-21}GeV$ (the bare mass of the axion $\theta$
from Eq.(\ref{pot})). One should notice that GSB mechanism predicts,
from the Hubble diagram of SNe Ia $alone$,
proportions of baryons and CDM ($\Omega_b$ and $\Omega_\theta$)
that are in agreement with WMAP constraints (see also \cite{awe}).
The GSB mechanism is based on two energy scales, one of ESB $\lambda_e$ and the other
of \textit{gravitational} symmetry breaking $\lambda_g$, replacing the SSB scale in the usual approach \cite{majoron,pngb}.
This scale $\lambda_g$ is related to the vev of the new complex scalar $\Psi$
through $<\Psi>=\Phi(a) \lambda_g$. We can now examine what are the cosmological constraints on $\lambda_e$ and $\lambda_g$ from the
Hubble diagram of type Ia supernovae. This is illustrated in Figure \ref{lgle}, where we plotted
the $95\%$ confidence contours of UNION SNe Ia data in the plane $(\lambda_g,\lambda_e)$. 
Cosmic acceleration
can be reproduced with the same accuracy as the concordance model $\Lambda$CDM from the large diagonal band in parameter space.
Actually, this band corresponds to the predominance of axion-like DM particle $\theta$ over
baryonic ordinary matter (see for instance the region $0.01<R_i<1$ that is also represented). Cosmic acceleration is therefore naturally
explained from a wide range of energy scales$\lambda_e$ and $\lambda_g$ in GSB, in fact roughly the same range that also explains the coincidence between
the cosmological abundances of baryons and CDM (cf. also \cite{awe}). Please note that the allowed region continues on the left of the diagram, we
have simply truncated it to $\lambda_g>10^{12}GeV$ for the sake of convenience.
In Figure \ref{lgle}, we have also represented some contours for the bare
mass of the axion-like DM particle $m_\theta$ obtained from the second derivative of the potential (\ref{pot}). It is important to notice that GSB
admits a much wider range of energy scales compatible with cosmic acceleration than the pNGB quintessence model \cite{pngb}.
In this model, the field $\theta$, which is  minimally coupled to gravity, acts as DE (and not as DM as is the case here) when the field slowly
rolls on the potential (\ref{pot}). This quintessence model explains
cosmic acceleration from a rather small range of ESB and SSB scales \cite{pngb}, as illustrated in 
Figure \ref{lgle}\footnote{The big box represents the range of scales considered in the original paper by Friemann and his collaborators \cite{pngb} while
the small one indicates the range allowed to account for cosmic acceleration.}.
The GSB
mechanism we have introduced here reproduces cosmic acceleration from a much wider range in parameter
space $(\lambda_g,\lambda_e)$. In particular, the ESB scale $\lambda_e$ does not need to be extremely small to account for cosmic
acceleration.

The GSB mechanism we have introduced here is very general since we did not specify
which global $U(1)$ symmetry could give the candidate for the axion (or pNGB) $\theta$.
Two famous possibilities are known: the Weinberg-Wilczek axion
with the Peccei-Quinn symmetry and the majoron with lepton number symmetry.
Although the Weinberg-Wilczek axion is of great interest to consider
with this mechanism, the phenomenological consequences on the physics
of the standard model are delicate to derive and the constraints are
likely to be very stringent. In addition, the energy scale at which cosmic
acceleration occurs, of order of the $meV$, is very far from the QCD scale.
That is why we prefer focusing here on the second famous possibility and
examine some possible perspectives of GSB mechanism applied to neutrino mass generation. Indeed, the vicinity of the neutrino mass discrepancy and the DE scale
has been long time conjectured to be related \cite{mavans,pngb}. In addition, only neutrinos
can couple to the complex scalar charged under $U(1)_{B-L}$ since
the physics of all other particles is $B-L$ invariant. Therefore, the action 
(\ref{gsb}), with no explicit coupling between the matter field $\psi_m$
and the non-minimally coupled $\Psi$, is completely justified by
this invariance.
\\
\\
The standard picture to give mass to the neutrinos is through the so-called see-saw mechanism.
Neutrino oscillations are explained by the fact that neutrino mass eigenstates
differ from the flavour eigenstates that interact.
Therefore, the lepton number symmetry is a global broken symmetry
since 
there exists right-handed neutrinos $\nu_R$ coupling to the
left-handed ones. 
The new scalar particle $\Psi$ that is charged under $U(1)_{B-L}$
couples directly to $\nu_R$ through a Majorana mass term $\mathcal{L}\approx
\Psi \bar\nu_R^c \nu_R$. The Majorana mass $m_{\nu_R}$ of $\nu_R$ is therefore
given by the vev of $\Psi$. While there
are no direct couplings with the other fields of the standard model
($B-L$ invariance), $\nu_R$ couples directly to the left-handed neutrinos $\nu_L$
through a Dirac mass term:  $m_D\bar\nu_R \nu_L$. 

\begin{figure}[h!t]
  \includegraphics[scale=0.4]{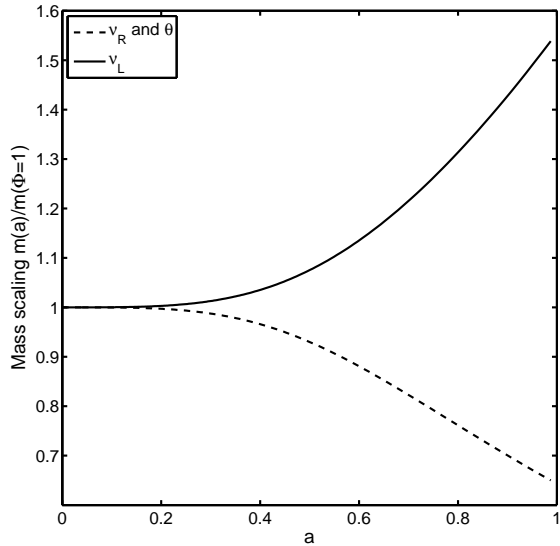}
  \caption{Cosmological evolution of the masses (in units of bare mass $\bar m=m(\Phi=1)$ for the GSB mechanism
applied to breaking of $U(1)_{B-L}$}
\label{figm}
\end{figure}

The see-saw mechanism constitutes in a mixing of both species
so that $m_{\nu_L}\approx m^2_D/m_{\nu_R}$ with the Dirac
mass typically of order of electroweak scale $m_D\approx 100GeV$. This see-saw
mechanism naturally explains the smallness of the left-handed neutrino mass
since the Majorana mass of the right-handed mass is expected to be very high, which
has to be achieved through the stabilisation of $<\Psi>$ with SSB.
In the GSB mechanism we have introduced, the vev of $\Psi$ (the field $\sigma$)
is stabilized through the cosmological convergence mechanism accompanying
the non-minimal gravitational coupling of $\Psi$.
\\
\\
We can now derive quantitative predictions for the (variation of) neutrino masses from the values of the parameters given above and obtained from the fit to Hubble diagram data.
From the energy scales used in Figure \ref{fig_acc}, the bare masses
of the neutrinos are given by $\bar m_{\nu_R}\approx 10^{14}GeV$ and 
$\bar m_{\nu_L}\approx 0.1 eV$. The reader should remember that a wide-range of
values of $\lambda_g$ allows accounting for cosmic acceleration in GSB mechanism,
so that many bare values of those neutrino masses are possible. In fact,
 cosmic acceleration does not constraint the bare values of these masses
but rather their variation, which is ruled by the cosmological evolution
of $\Phi$. Figure \ref{figm} gives the variation of the neutrino 
and majoron $\theta$ masses\footnote{The masses scales as
$m_{\nu_R}\sim m_\theta\sim \Phi$ and $m_{\nu_L}\sim\Phi^{-1}$.}
on cosmological scales predicted by the GSB applied to cosmic acceleration.
A mass variation of about 30\% of the Majoron between the CMB epoch and
today sufficies to account for cosmic acceleration. As a consequence, provided
GSB is the real mechanism that breaks the lepton number symmetry,
the right-handed neutrinos masses decreases of about 30\%
while the left-handed neutrinos masses increases of about 60\%,
in the same time range. 
This prediction is of first importance for cosmic
structure formation since the DM particles
(here both species neutrinos and majorons) could potentially change
from cold to warm DM depending on their bare masses. The other important
impact is the large-scale modification of gravitation induced by the abundance
of DM particles on galactic scales.
\section{Conclusions}
The question whether the WEP, or equivalently the universality of free-fall, also applies to DM is strongly justified,
particularly if DM originates from some symmetry breaking scenario. This possibility opens new fascinating 
perspectives for cosmology since such violation of the WEP yields important impacts
on cosmic expansion, large-scale structure formation and galactic dynamics. In addition, since the WEP has to be relaxed
if it is proved that it does not apply to DM, the SEP has also to be revised. This opens the way to modifications of gravity
induced by the abundance of abnormally weighting DM.\\
\\
In this paper, we have introduced a new mechanism allowing to break global
$U(1)$ symmetries through a cosmological relaxation described by the AWE tensor-scalar theory of gravitation \cite{awe}.
In this new picture, DM is constituted by axion-like particles which are the pseudo-Nambu-Goldstone boson emerging
from the ESB. The mass of these axion-like particles is stabilised by a gravitational process
in which the strength of gravity is also evolving (non-minimal gravitational coupling). 
This results in a non FLRW cosmic expansion that can produce cosmic acceleration without negative pressures, when
the DM mass is varying. We have used the Hubble diagram of far-away SNe Ia to constraints the two
energy scales  of the mechanism: $\lambda_g$ the GSB scale (replacing the usual SSB scale) and $\lambda_e$ the ESB scale.
A wide range of scales is allowed to account for the enigmatic cosmic acceleration, which also turn out
to be the range explaining the observed cosmological abundance of DM and baryons.
We have then applied this new GSB mechanism to the breaking of the global lepton number symmetry $U(1)_{B-L}$,
which is the standard picture to explain non-vanishing neutrino masses. From the choice of scales $\lambda_g, \lambda_e$
that are both in agreement with cosmic and particle physics constraints, the model not only predicts the
mass amplitudes of neutrinos but also the cosmological variation of these masses between CMB epoch and today.
This mass variation of DM and neutrinos and the associated modified gravity effects open interesting perspectives for cosmic structure formation and galactic dynamics. Constraints from neutrino physics should also give the way to falsify this prediction of GSB.\\
\\
In conclusion, the key questions of DM and DE in cosmology lead us to introduce here a new possible bridge between two pillars of modern physics:
the equivalence principle and symmetry breaking. Once again, investigating the invisible universe seems to be a path to
a unified description of the infinitely large and the infinitely small.


\begin{theacknowledgments}
One of the authors (A.F.) is grateful to Pr. 
E. Bertschinger, Pr. A. Guth from MIT and Pr. J.-M. G\'erard from UCLouvain for insightful discussions that helped initiate the fundamental idea of this work. Numerical simulations were made on the local computing resources
at Unit\'e de Syst\`emes Dynamiques (FUNDP, Belgium).
\end{theacknowledgments}



\bibliographystyle{aipproc}   

\begin{thebibliography}{9}
 \bibitem{axion}
S. Weinberg, Phys. Rev. Letters 40(1978), p. 223;\\
F. Wilczek, Phys. Rev. Letters 40(1978), p. 279
\bibitem{PQ}
R. D. Peccei, H. R. Quinn, Physical Review Letters, 38(1977) p. 1440,
 Physical Review, D16 (1977) p. 1791-1797.
\bibitem{majoron}
Y. Chikashige, R.N. Mohapatra, R.D. Peccei, Phys.Lett. B98, 265 (1981).
\bibitem{kolb} E.W. Kolb, M.S. Turner, \textit{The Early Universe}, Addison-Wesley, 1994.
\bibitem{farrar} G.R. Farrar \& P.J.E. Peebles, Astrophys. J. \textbf{604} 1-11 (2004).
\bibitem{massd} C.T. Will \& G.C. Ross, Nucl. Phys. B 311, 253 (1988);\\
J. Ellis, S. Kalara, K.A. Olive \& C. Wetterich, Phys. Lett. B 228, 264 (1989);\\
C. Wetterich, Astron. Astrophys. 301, 321 (1995);\\
G.W. Anderson \& S.M. Caroll, astro-ph/9711288;\\
G. Huey, P.J. Steinhardt, B.A. Ovrut \& D. Waldram, Phys. Lett. B 476, 379 (2000);\\
D. F. Mota, J. D. Barrow, Mon. Not. Roy. Astron. Soc. 349, 291 (2004), astro-ph/0309273\\
D. F. Mota, J. D. Barrow, Phys.Lett.B 581 141-146 (2004), astro-ph/0306047\\
A. Ringwald \& L. Schrempp, JCAP 0610, 012 (2006)\\
A. W. Brookfield  et al., PRD 73, 083515 (2006)\\
\bibitem{mavans}
R. Fardon, A. E. Nelson and N. Weiner, JCAP. 0410, 005 (2004)\\
R.D. Peccei, Phys. Rev. D 71 023527 (2005) ;\\
C. Wetterich, Phys.Lett.B 655 201 (2007) ;\\
 L. Amendola, M. Baldi, C. Wetterich, Phys. Rev. D 78, 023015 (2008)
\bibitem{dmde}
L. Amendola, Phys.Rev. D62 (2000) 043511, \\ 
S. Das, P.-S. Corasaniti \& J. Khoury, Phys.Rev. D73 (2006) 083509
\bibitem{su} Y. Su et al. Phys. Rev. D 50, 3614 - 3636 (1994)
\bibitem{kamionkowski} M. Kesden \& M. Kamionkowski, Phys.Rev. D74 (2006) 083007;
Phys.Rev.Lett. 97 (2006) 131303.
\bibitem{awe} J.-M. Alimi \& A. F\"uzfa, JCAP 09, 014 (2008)\\
A. F\"uzfa \& J.-M. Alimi, Phys. Rev. Lett. \textbf{97} 061301 (2006)\\
A. F\"uzfa \& J.-M. Alimi, Phys. Rev. D \textbf{75} 123007 (2007)\\
J.-M. Alimi \& A. F\"uzfa, Int.J. Mod. Phys. D 16, 2587 - 2592 (2007)\\
A. F\"uzfa \& J.-M. Alimi, Phys. Rev. D \textbf{73} 023520 (2006).
\bibitem{will} C.M. Will, Liv. Rev. Rel. \textbf{9}, 3 (2006).
\bibitem{mantry}
S. M. Carroll, S. Mantry, M. J. Ramsey-Musolf, C. W. Stubbs,
Phys.Rev.Lett.103:011301 (2009)\\
S. M. Carroll, S. Mantry, M. J. Ramsey-Musolf, arXiv:0902.4461
\bibitem{convts} T. Damour, G.W. Gibbons \& C. Gundlach, Phys. Rev. Lett. \textbf{64}, 123 (1990)\\
T. Damour and K. Nordtvedt, Phys. Rev. D48 (8),
  3436-3450 (1993);\\
T. Damour and K. Nordtvedt, Phys. Rev. Lett. 70 (15),
  2217-2219 (1993)
\bibitem{serna}
A. Serna and J.-M. Alimi, Phys.Rev. D53, 3074 (1996);\\
  A. Serna, J.-M. Alimi and A. Navarro, Class.Quant.Grav. 19 (2002) 857-874
  \bibitem{union} Kowalski, M., et al. 2008, Astrophys.J.686:749-778,2008
  \bibitem{pngb}J. Frieman, C. T. Hill, A. Stebbins, I. Waga, Phys. Rev. Lett. 75, 2077 - 2080 (1995);\\
  E. Di Pietro, J.-F. Claeskens, Mon.Not.Roy.Astron.Soc. 341 (2003) 1299 ;\\
  L. Amendola, R. Barbieri, PLB (2006);\\
  E.J. Copeland, M. Sami \& S. Tsujikawa, Int.J.Mod.Phys. D15 1753-1936 (2006).
  \bibitem{das} S. Das, N. Weiner, astro-ph/0611353
\bibitem{cplxdmde}
R. Mainini, S. Bonometto, Phys. Rev. Lett. 93, 121301 (2004)
A. Arbey, Phys. Rev. D 74, 043516 (2006)

\bibitem{alimi} J.-M. Alimi, A. F\"uzfa, ``\textit{The Abnormally Weighting Energy Hypothesis: The origin of the cosmic acceleration}'', in this volume.

\end{thebibliography}

\IfFileExists{\jobname.bbl}{}
 {\typeout{}
  \typeout{******************************************}
  \typeout{** Please run "bibtex \jobname" to optain}
  \typeout{** the bibliography and then re-run LaTeX}
  \typeout{** twice to fix the references!}
  \typeout{******************************************}
  \typeout{}
 }

\end{document}